\documentclass{article}        
\usepackage{graphicx}
\usepackage{amsmath,amssymb,amsfonts}
\usepackage[english]{babel}
\usepackage[bookmarks=false]{hyperref}
\begin{document}
\title{Study of the ferromagnetic quantum phase transition in Ce$_{3-x}$Mg$_x$Co$_{9}$}
\maketitle
\author{Tej N. Lamichhane$^{1,2}$},
\author{Valentin Taufour$^{1,3}$},
\author{Andriy Palasyuk$^{1,2}$},
\author{Sergey L. Bud'ko$^{1,2}$} and 
\author{Paul C. Canfield$^{1,2}$}

$^1$: Ames Laboratory, U.S. DOE, Ames, Iowa 50011, USA \\
$^2$:Department of Physics and Astronomy, Iowa State University, Ames, Iowa 50011, USA\\
$^3$: Department of Physics, University of California Davis, Davis, California 95616, USA

\begin{abstract}
The Ce$_{3-x}$Mg$_x$Co$_{9}$ system evolves from a Pauli paramagnetic ground state for $x = 0$ to a ferromagnetic ground state for $x\approx0.80$ in single phase, polycrystalline samples [Phys. Rev. Applied 9, 024023 (2018)]. In order to better understand this behavior, single crystalline samples of Ce$_{3-x}$Mg$_x$Co$_{9}$ for \textit{x} = 0.01, 0.16, 0.24, 0.35, 0.43 and 0.50 were grown using the flux growth technique, and electrical transport and magnetic properties were studied. The \textit{T}$_C$-\textit{x}  phase diagram we infer shows that the system has a quantum phase transition near \textit{x} = 0.35, transforming to a ferromagnetic ground state.

\end{abstract}

\maketitle

\section{Introduction}

Fragile magnetic systems~\cite{Canfield_2016} tuned via chemical substitution  can lead to quantum criticality, or more generally, a quantum phase transition at 0 K. Here we investigate a Pauli paramagnetic to ferromagnetic quantum phase transition in Ce$_{3-x}$Mg$_x$Co$_9$~\cite{TejCeCo32018,PandeyCeCo32018}.

\begin{figure}[!h]
\centering
\includegraphics[scale =0.35]{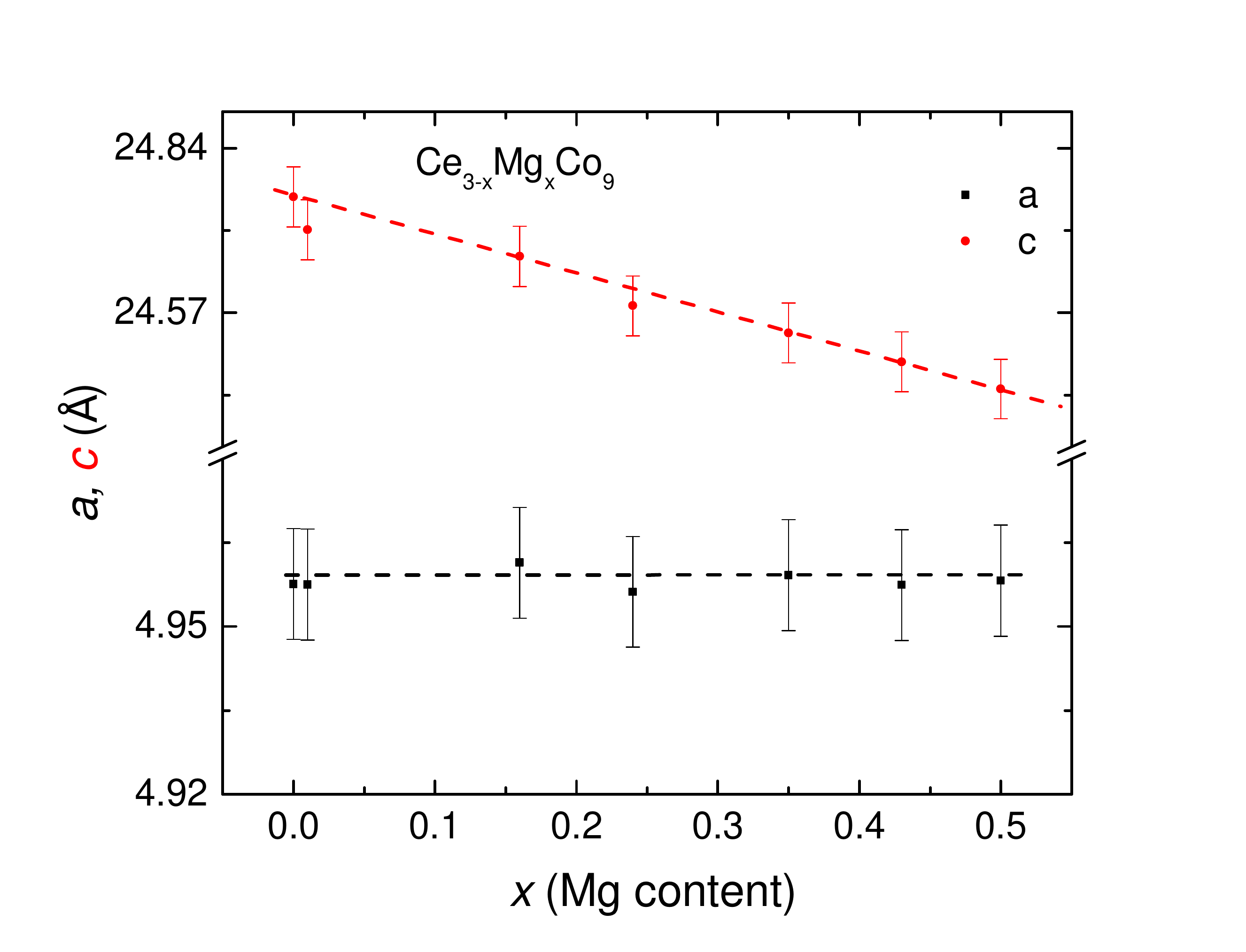}
\caption{Variation of lattice parameters of Ce$_{3-x}$Mg$_x$Co$_9$ with Mg content \textit{x} inferred from EDS. The uncertainty of the lattice parameters values is less than 0.2\%. The dashed lines are guides to the eye.}
\label{Latticeparametersfinal}
\end{figure}

Despite being 75 \% atomic Co, pure CeCo$_3$ has a Pauli paramagnetic, low temperature ground state. In polycrystalline Ce$_{3-x}$Mg$_x$Co$_9$, transformation of a Pauli paramagnetic CeCo$_3$ into a ferromagnetic phase was studied in our earlier work~\cite{TejCeCo32018}. For single phase samples, ferromagnetism was observed for $x \geq 0.80$. In mixed phase samples, signs of magnetic order could be found for $x< 0.80$ but in such samples, magnetism could be influenced by several factors such as defects~\cite{Xing2011}, stress~\cite{Shimuzu2015}, impurities~\cite{Zhao2002} etc. To elucidate the phase transformation in a much cleaner way, study of magnetism using single phase, single crystalline samples is always a better idea. We investigated the magnetic properties, electrical transport and specific heat capacity of selected compositions around x$\sim$0.35 of single crystalline  Ce$_{3-x}$Mg$_x$Co$_9$ samples. 

We find a quantum phase transition as Ce$_{3-x}$Mg$_x$Co$_9$ changes from  a Pauli paramagnetic state for 0$\leq x < 0.35$ to a ferromagnetic state for $0.35< x < 1.4$. No additional magnetic phases were found in the vicinity of the quantum phase transition composition.

\section{Experimental details}
Single crystalline Ce$_{3-x}$Mg$_x$Co$_9$ samples for $x \leq 0.5$ were synthesized in 3-cap Ta crucibles~\cite{Canfield2001JCG} similar to the previous report~\cite{TejCeCo32018} by maintaining the Ce$+$Mg to Co ratio as 30:70 i.e. (Ce$_{1-x}$Mg$_x$)$_{30}$Co$_{70}$ with nominal \textit{x} = 0.05, 0.10, 0.15, 0.20, 0.25 and 0.30 as listed in Table~\ref{tbl:composition}.

\begin{table}[!h]
\begin{center}
\caption{Loaded nominal, EDS and Rietveld refined composition of Ce$_{3-x}$Mg$_x$Co$_9$  samples for crystal growth. The uncertainty in Mg content is given in parenthesis which was obtained as a standard deviation of EDS measurement. The average uncertainty in Mg concentration in the Rietveld refinement is $\leq \pm 0.05$.}
\begin{tabular}{| p{2.50cm} | p{3.3cm} | p{2.5cm}|}
\hline
Loaded~nominal\newline composition  &  EDS composition & Rietveld refined composition  \\ \hline
Ce$_{30}$Co$_{70}$ & --- & CeCo$_3$   \\ \hline
Ce$_{28.5}$Mg$_{1.5}$Co$_{70}$ &  Ce$_{2.99}$Mg$_{0.01(0.01)}$Co$_9$ &   Ce$_{2.95}$Mg$_{0.05}$Co$_9$ \\ \hline
Ce$_{27}$Mg$_{3}$Co$_{70}$  &  Ce$_{2.84}$Mg$_{0.16(0.02)}$Co$_9$ &  Ce$_{2.87}$Mg$_{0.13}$Co$_9$\\ \hline
Ce$_{25.5}$Mg$_{4.5}$Co$_{70}$  &  Ce$_{2.76}$Mg$_{0.24(0.02)}$Co$_9$ &   Ce$_{2.75}$Mg$_{0.25}$Co$_9$ \\ \hline
Ce$_{24}$Mg$_{6}$Co$_{70}$  &  Ce$_{2.65}$Mg$_{0.35(0.015)}$Co$_9$ &   Ce$_{2.68}$Mg$_{0.32}$Co$_9$ \\ \hline
Ce$_{22.5}$Mg$_{7.5}$Co$_{70}$ &  Ce$_{2.57}$Mg$_{0.43(0.015)}$Co$_9$ &   Ce$_{2.57}$Mg$_{0.43}$Co$_9$ \\ \hline
Ce$_{21}$Mg$_{9}$Co$_{70}$ &  Ce$_{2.50}$Mg$_{0.50(0.02)}$Co$_9$ &   Ce$_{2.49}$Mg$_{0.51}$Co$_9$ \\ \hline
\end{tabular}
\label{tbl:composition}
\end{center}
\end{table}

\begin{figure}[!h]
\centering
\includegraphics[scale =0.35]{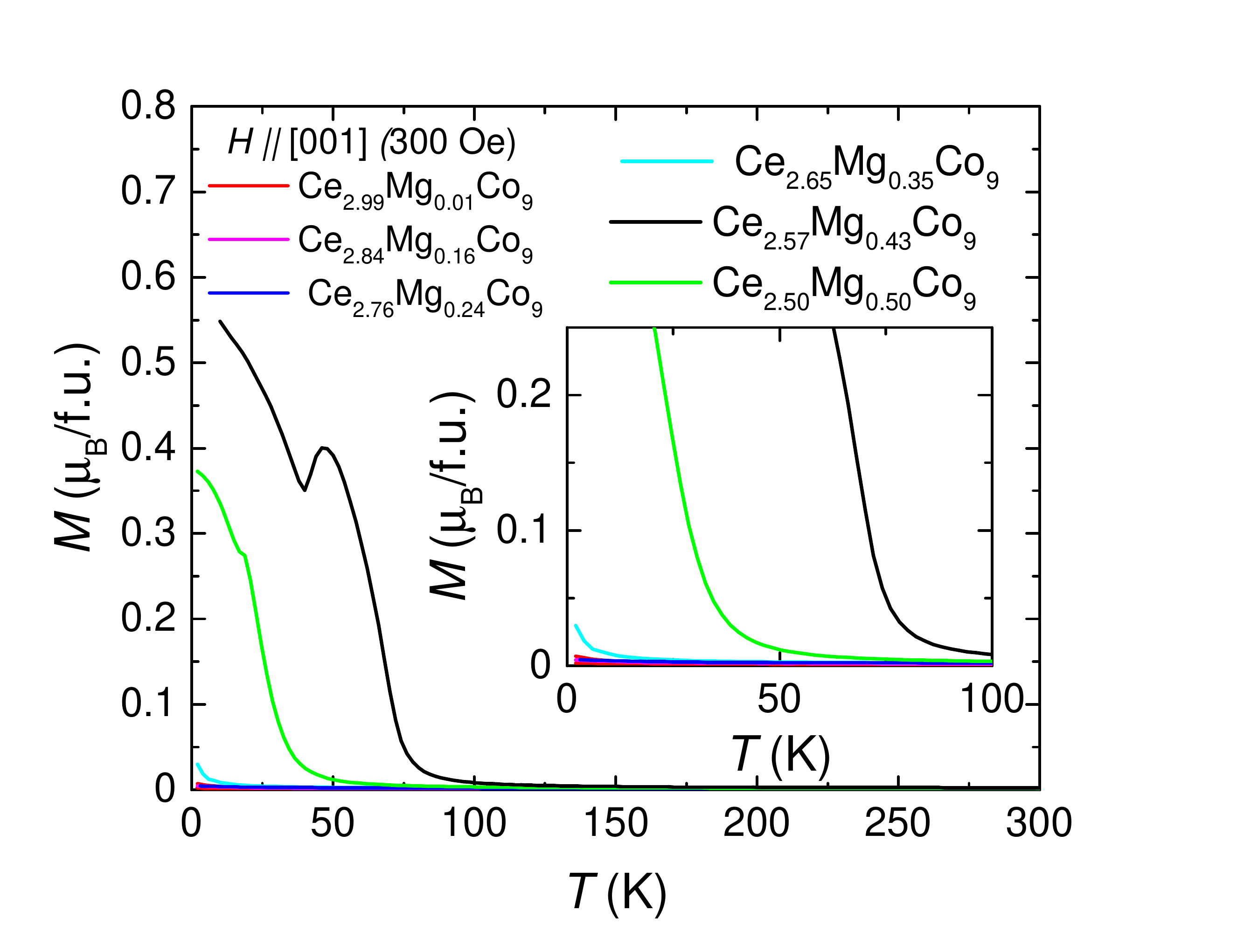}
\caption{ZFC temperature  dependent magnetization \textit{M(T)} data of various Ce$_{3-x}$Mg$_x$Co$_9$ samples at 300 Oe applied field. The inset shows the enlargement around the upturns to highlight the ferromagnetism development with higher concentrations of Mg.}
\label{nonmag}
\end{figure}

\begin{figure}[!h]
\centering
\includegraphics[scale =0.35]{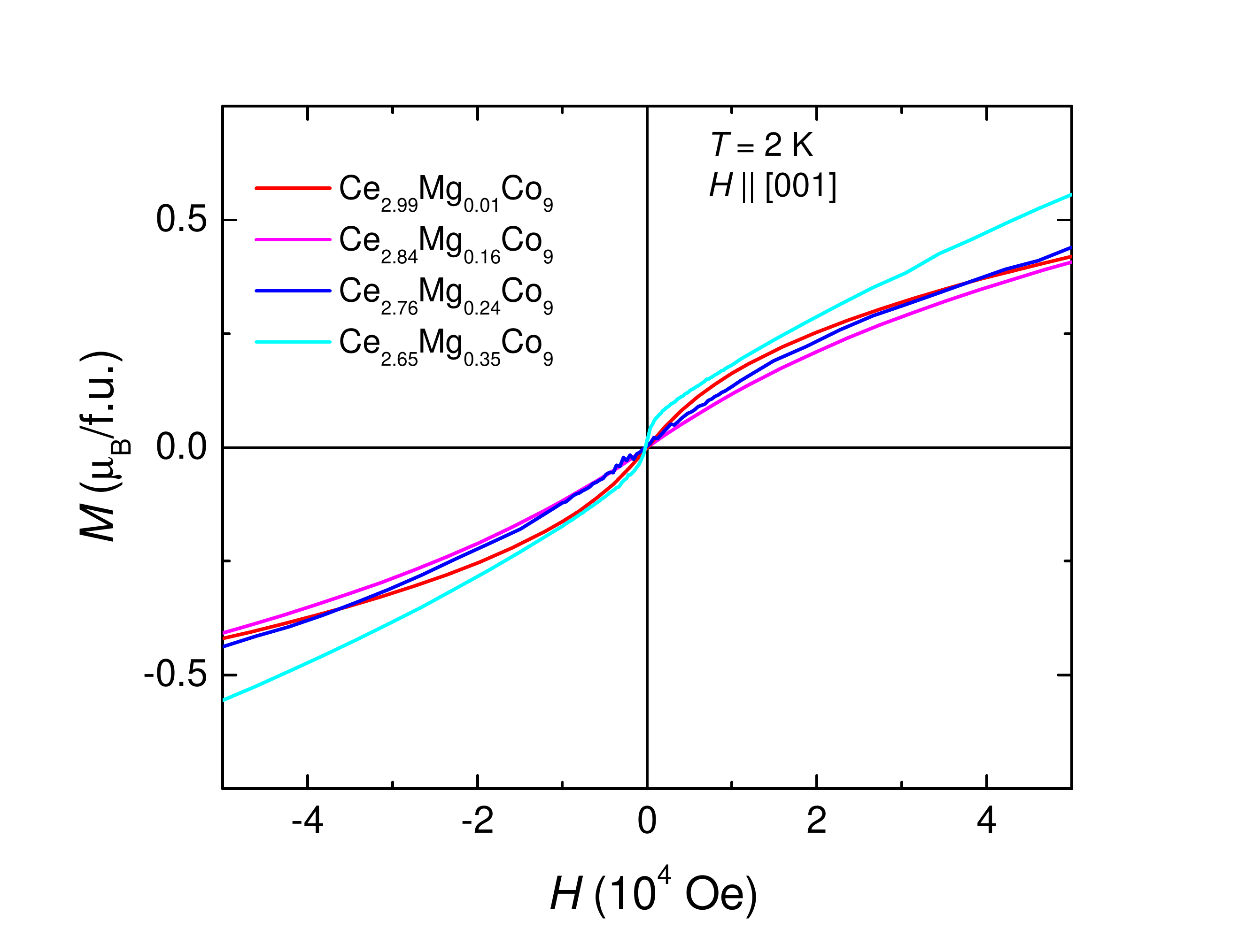}
\caption{Field dependent magnetization of various non-ferromagnetic Ce$_{3-x}$Mg$_x$Co$_9$ samples at 2 K parallel to \textit{c} axis.}
\label{MH-nonmag}
\end{figure}

\begin{figure}[!h]
\centering
\includegraphics[scale =0.35]{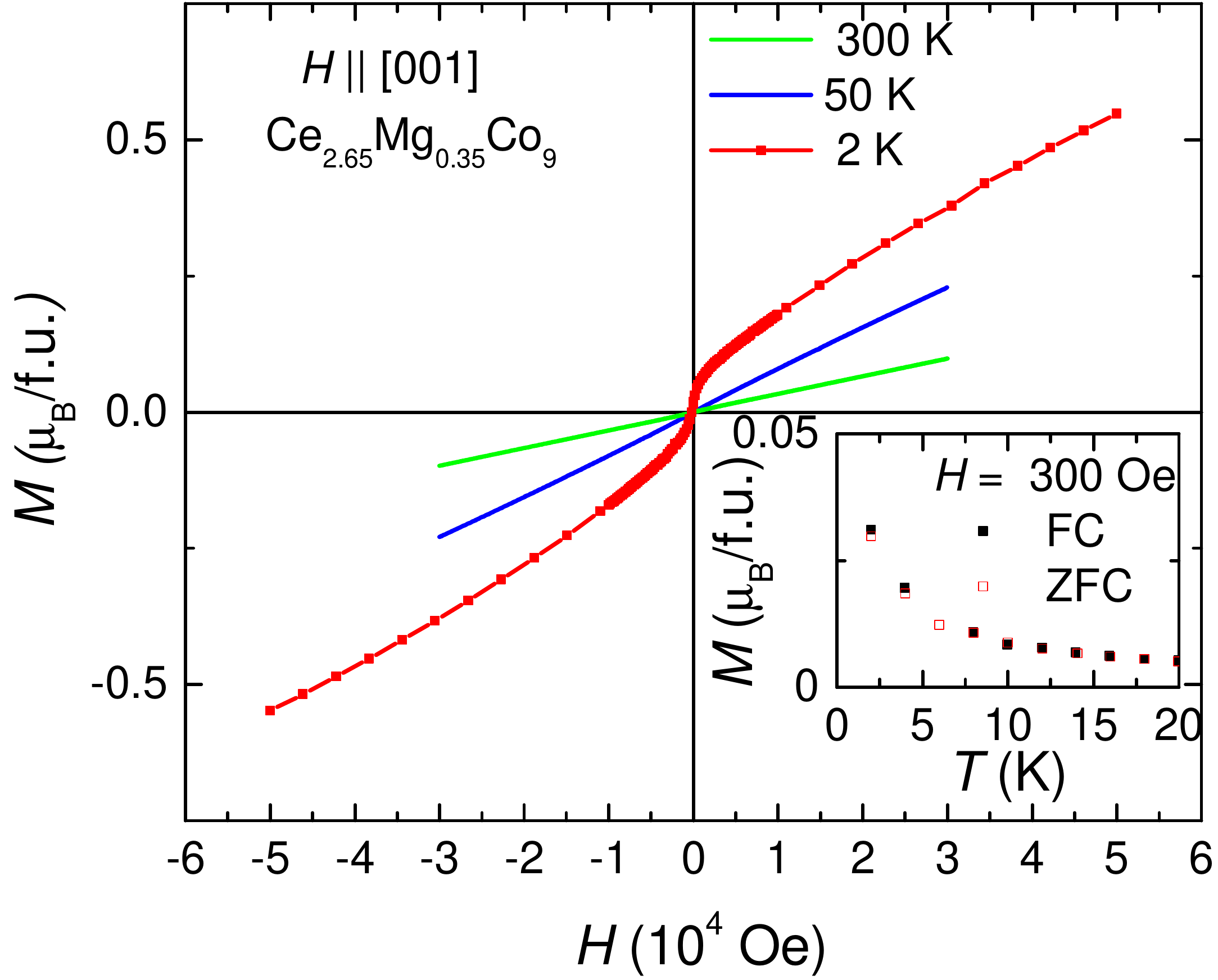}
\caption{Field dependent magnetization of Ce$_{2.65}$Mg$_{0.35}$Co$_9$ at various temperature to demonstrate only base temperature has some Brillouin type saturation magnetization. The lower inset shows reversible nature of the ZFC and FC \textit{M(T)} data.}
\label{TL922-MH}
\end{figure}

\begin{figure}[!h]
\centering
\includegraphics[scale =0.35]{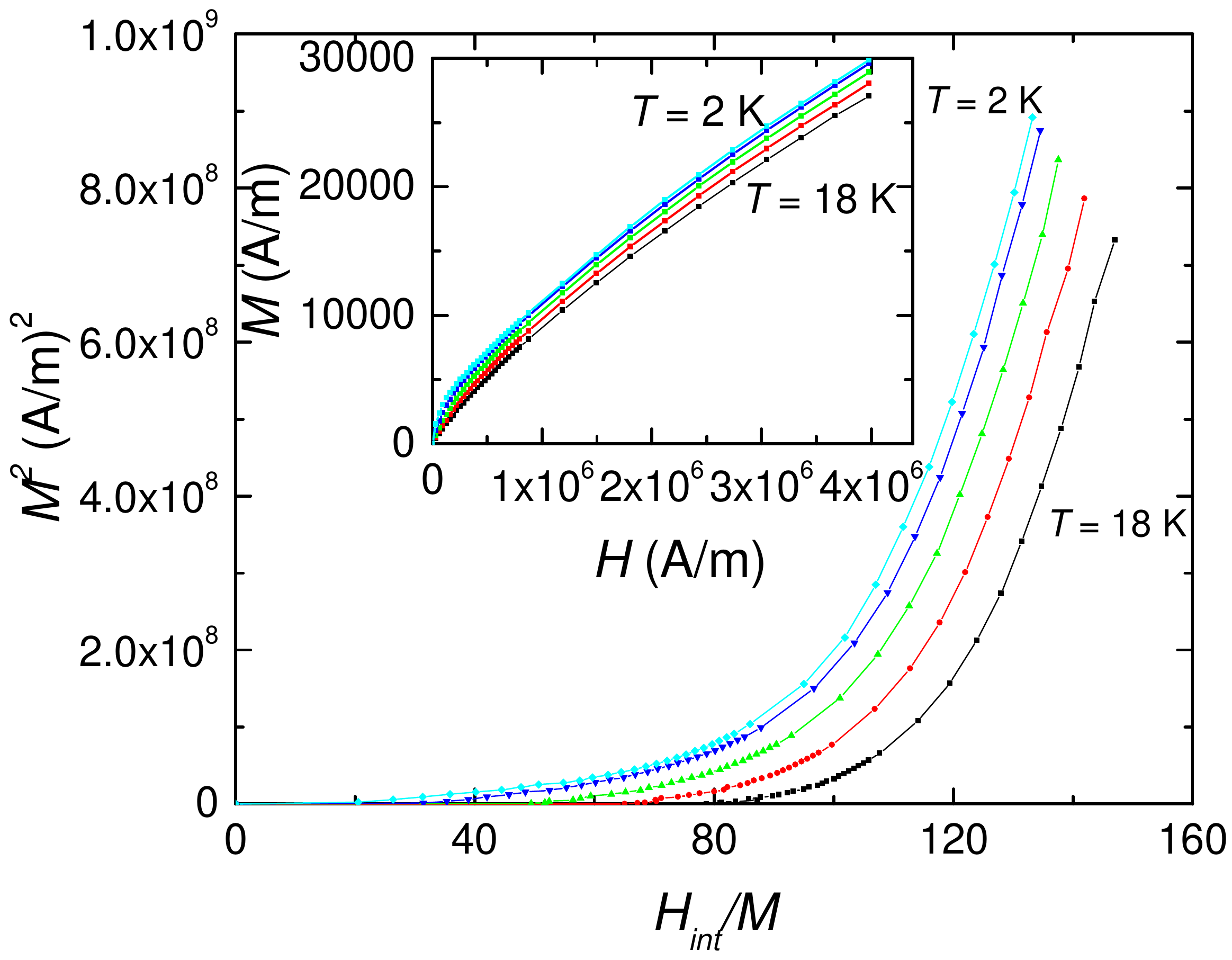}
\caption{The Arrott plot of Ce$_{2.65}$Mg$_{0.35}$Co$_9$  for $2~{\rm K} \leq T \leq 18$ K at a step of 4 K. The Curie temperature is suggested to be lower than 2 K. The inset shows the corresponding \textit{M(H)} data.}
\label{TL922Arrott}
\end{figure}

\begin{figure}[!h]
\centering
\includegraphics[scale =0.35]{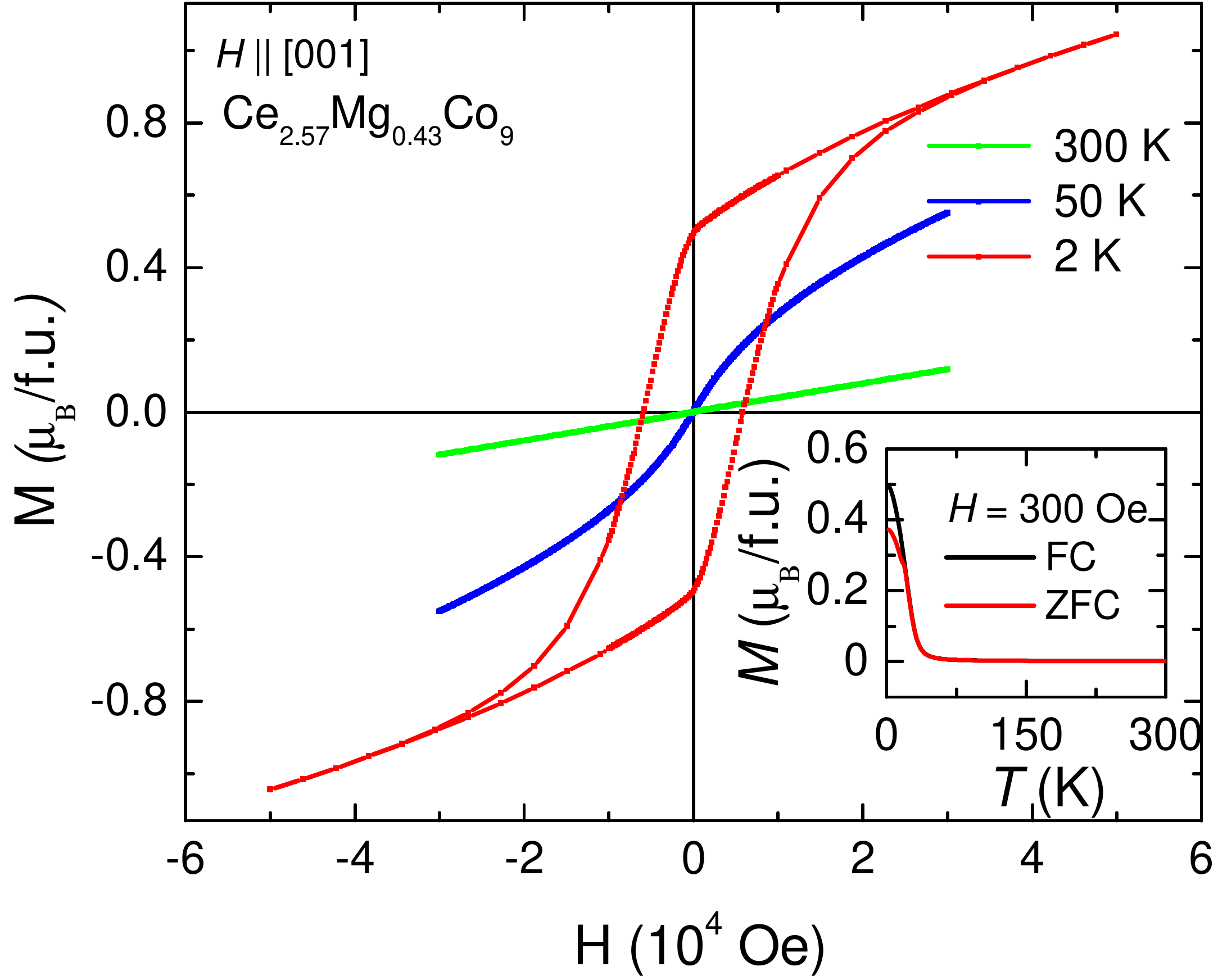}
\caption{Field dependent magnetization of Ce$_{2.57}$Mg$_{0.43}$Co$_9$ at various temperatures. The hystersis loop observed for 2 K data clearly indicates a ferromagnetic state and even suggests that Ce$_{3-x}$Mg$_x$Co$_9$ system may have good pinning properties for possible permanent magnet applications. The lower inset shows the \textit{M(T)} along with bifurcation of ZFC and FC \textit{M(T)} data consistent with observed low temperature hysteresis loop.}
\label{MH-TL923}
\end{figure}

\begin{figure}[!h]
\centering
\includegraphics[scale =0.7]{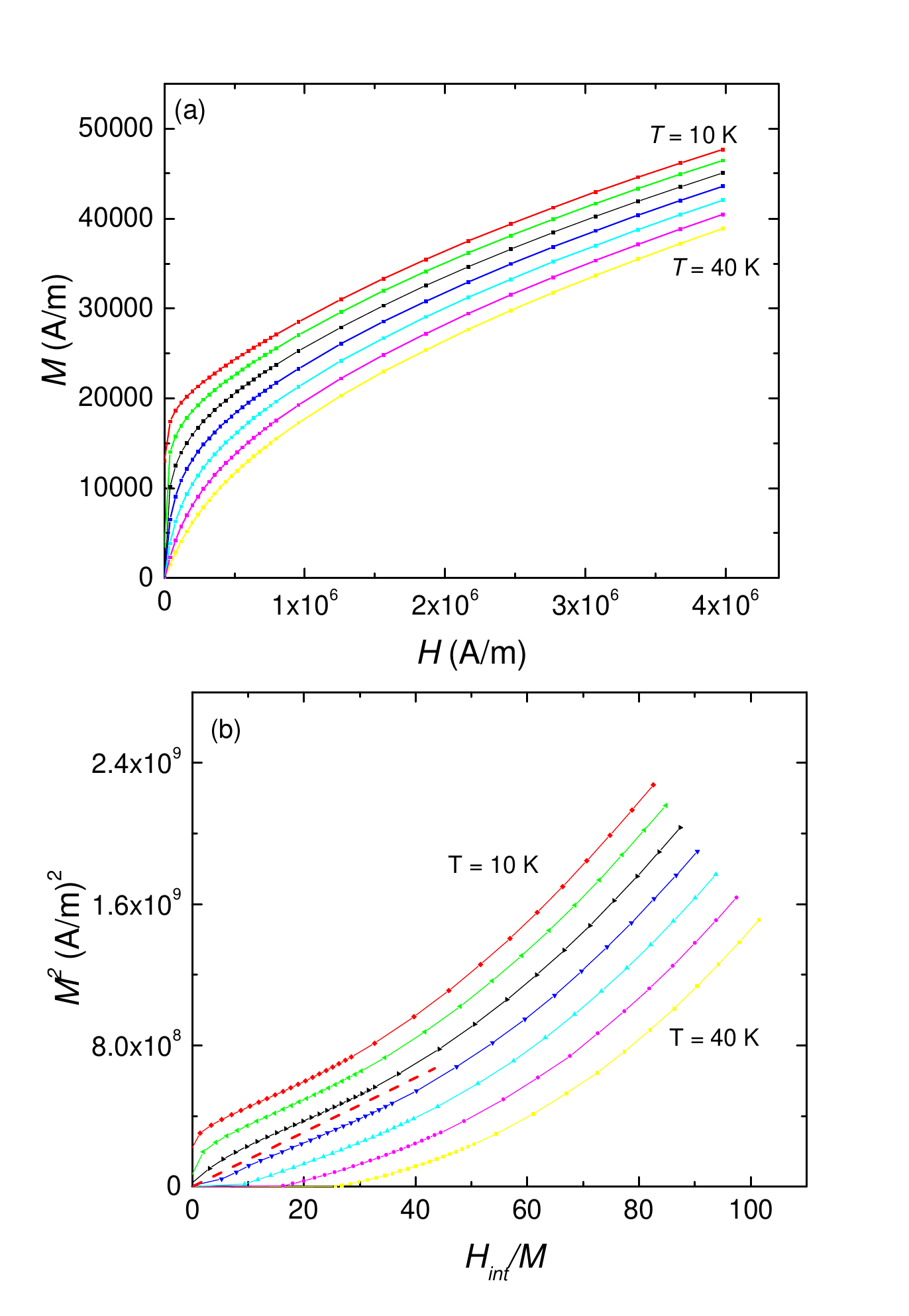}
\caption{(a) \textit{M(H)} data for Ce$_{2.57}$Mg$_{0.43}$Co$_9$  sample around the Curie temperature for $10~{\rm K}~\leq T \leq 40$ K at a step of 5 K. (b) Corresponding Arrott plot to determine the Curie temperature which is found to be around 25 K.}
\label{TL923Arrott}
\end{figure}

\begin{figure}[!h]
\centering
\includegraphics[scale =0.35]{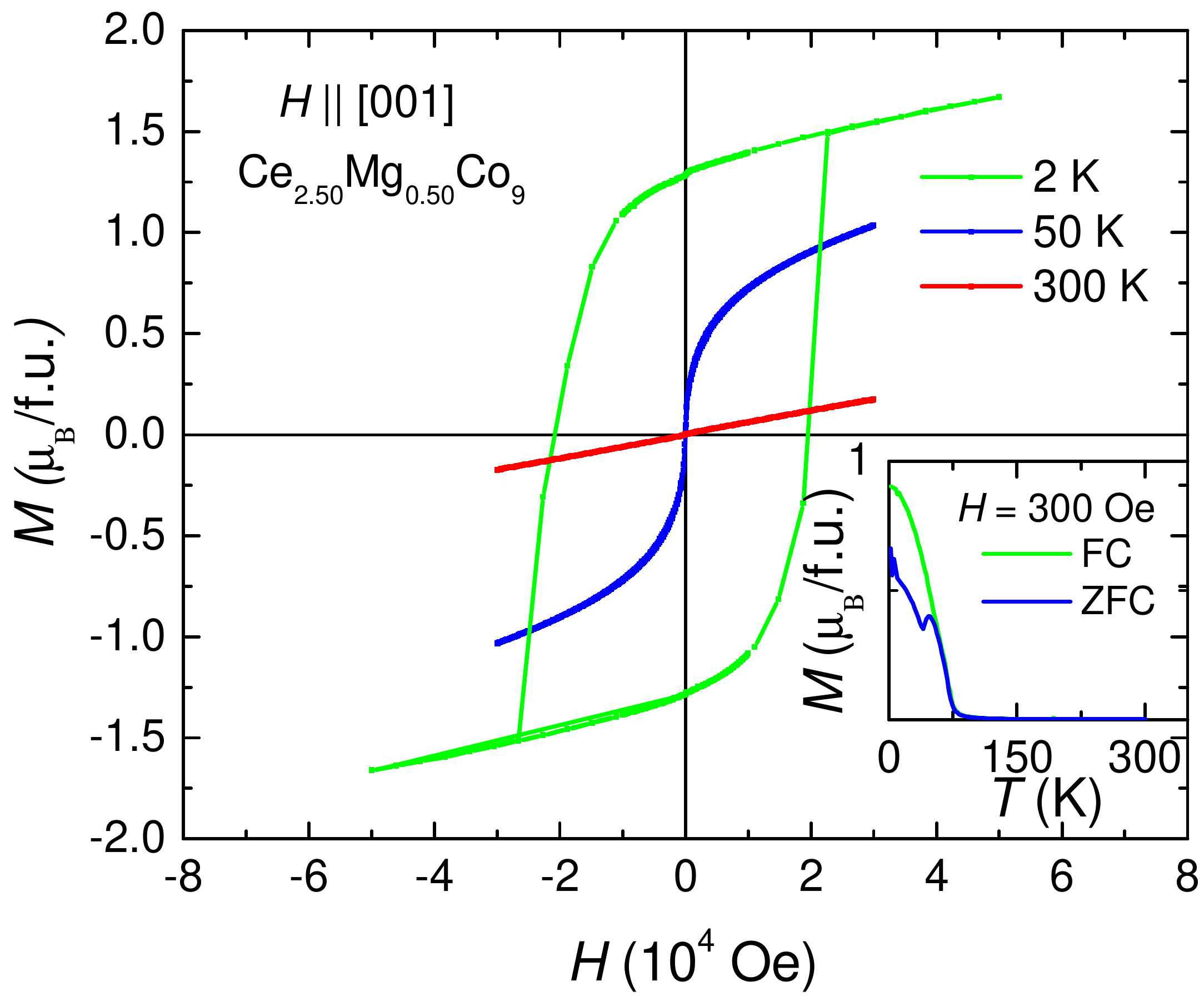}
\caption{Field dependent magnetization of Ce$_{2.50}$Mg$_{0.50}$Co$_9$ at various temperatures (Note: hysteresis at 2 K is larger than that shown for \textit{x} = 0.43 as shown in Fig.~\ref{MH-TL923}). The lower inset shows the \textit{M(T)} along with bifurcation of ZFC and FC \textit{M(T)} data consistent with observed low temperature hysteresis loop.}
\label{MH-TL924}
\end{figure}

\begin{figure}[!h]
\centering
\includegraphics[scale =0.35]{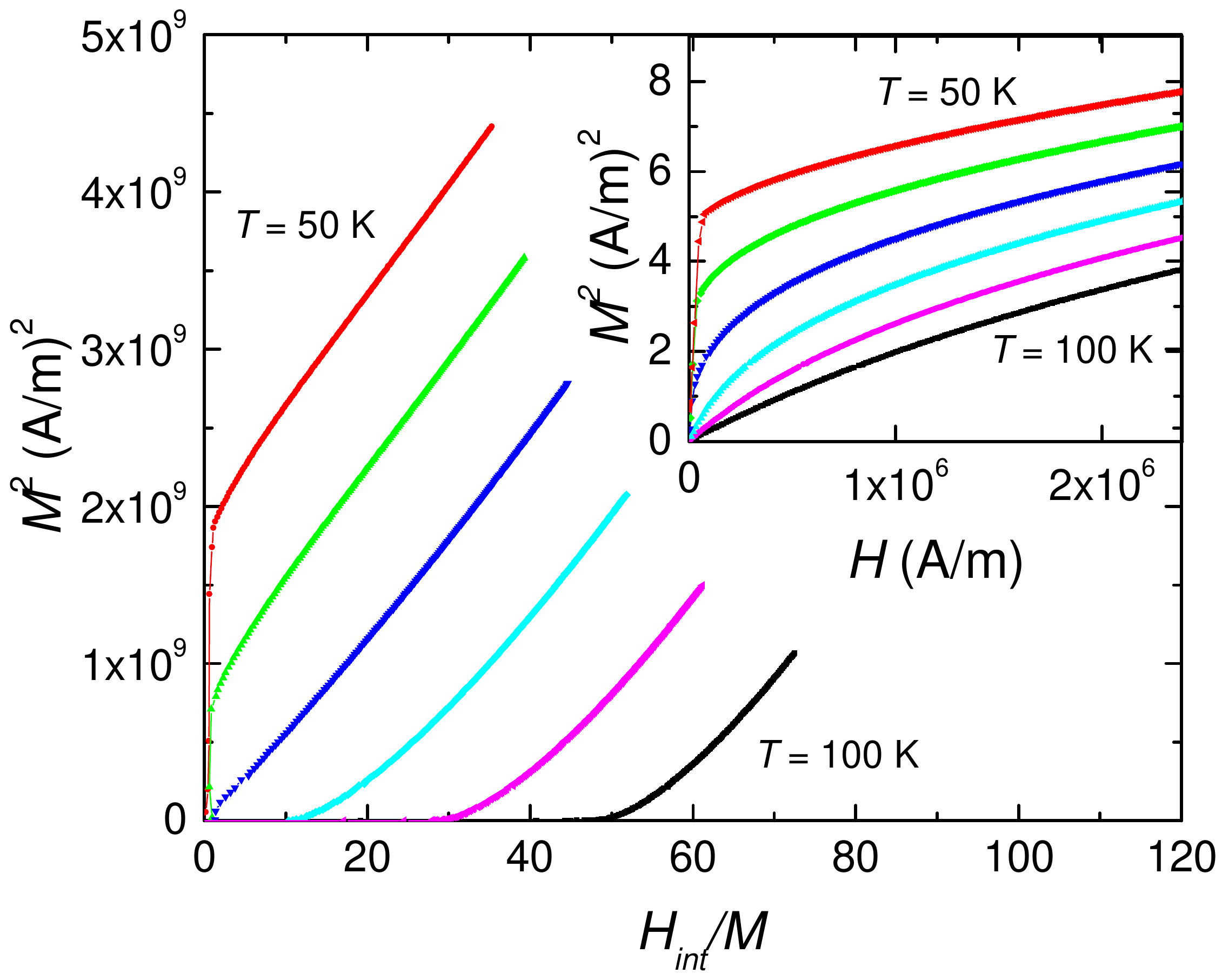}
\caption{The Arrott plot of Ce$_{2.50}$Mg$_{0.50}$Co$_9$  for $50~{\rm K}~\leq T~100$ K at a step of 10 K. The Curie temperature is determined to be around 70 K. The inset shows the corresponding \textit{M(H)} data.}
\label{TL924Arrott}
\end{figure}

\begin{figure}[!h]
\centering
\includegraphics[scale =0.35]{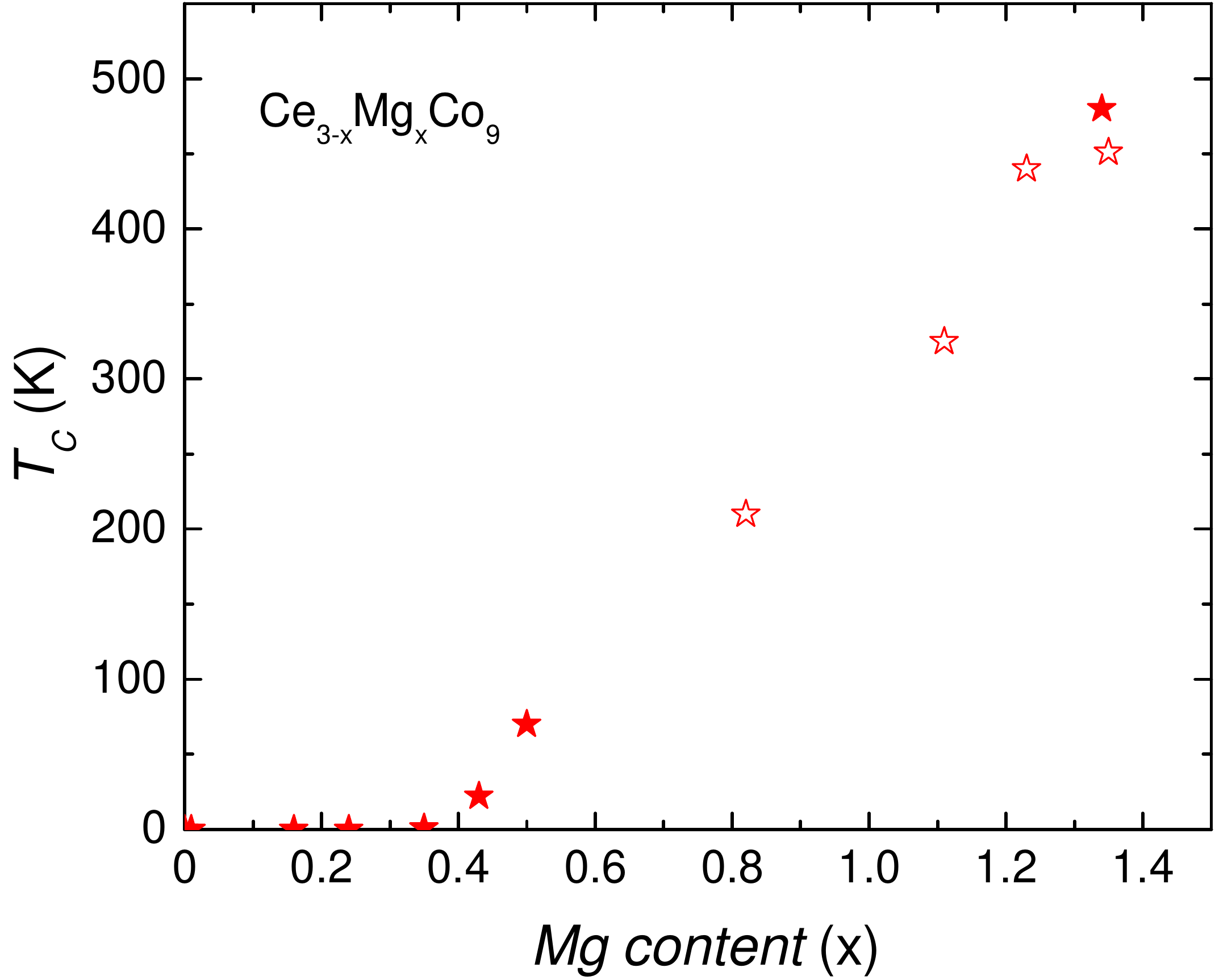}
\caption{\textit{T}$_C$ - \textit{x} phase diagram of Ce$_{3-x}$Mg$_x$Co$_9$ samples. Above $x \sim 0.35$, the Curie temperature is approximately linearly proportional to Mg content. The solid stars represents the Curie temperatures measured via Arrott plot analysis on single crystalline samples. The hollow stars represents Curie temperatures measured on polycrystalline samples~\cite{TejCeCo32018}.}
\label{xTC-Phasediagram}
\end{figure}

The selected amount of Ce-Co-Mg elements were sealed under almost one atmospheric pressure of Ar at room temperature inside a 3-cap Ta crucible. The Ta crucibles were protected under a partial pressure of  Ar inside an amorphous silica jacket with quartz wool buffers on the top and the bottom of the crucible to protect silica ampoule from the consequences of differential thermal expansion during temperature ramping and mechanical shock during the decanting process. The silica ampoule was heated to 900~$^\circ$C over 3 h and held there for 3 h to allow the initial reaction of elements and potentially avoid the excessive boiling of Mg at higher temperature. Then the growth was heated to 1200~$^\circ$C over 3 h and held there for 10 h to form a homogeneous liquid. Finally, the growth was slowly cooled down to 1040~$^\circ$C over $\sim$100 h for nominal Mg content \textit{x} = 0.05 - 0.15 and centrifuged to separate the crystals from the flux. Similarly, growths with \textit{x} = 0.20 - 0.30  were cooled down to 1070~$^\circ$C over $\sim$100 h and crystals were separated. The single crystalline samples had rhombohedral plate-like morphology with several millimeters in length and width and approximately 1 - 2 mm in thickness. The easy axis of magnetization is perpendicular to the plate ([001] direction)~\cite{TejCeCo32018}.

Crystals from all the batches were characterized using powder X-ray diffraction (XRD) and Scanning Electron Microscopy (SEM).  For XRD, a finely ground powder was spread over the zero background silicon wafer and held in place with a thin film of Dow Corning high vacuum grease. Powder XRD data were collected using Rigaku Miniflex II diffractometer within 2$\theta$ range of 5 - 100$^\circ$ using a step size of 0.01 degree and a dwell time of 3 seconds. For SEM, single crystalline samples were mounted in epoxy and finely polished perpendicular to the plane of the plate to determine the composition. The nominal composition, average Energy Dispersive Spectroscopy (EDS) composition, and Rietveld refined composition are presented in Table I. For nominal Mg  \textit{x } = 0.05, 0.10 and 0.15 samples, traces of CeCo$_2$ impurities were visible in the cross sectional view of SEM images (not shown here). The EDS and Rietveld refined composition more or less agree each other except for nominal 5\% Mg doped sample. Hereafter, all the compositions are  EDS inferred in this paper. 

Magnetic properties were measured in a Quantum Design (QD) Magnetic Property Measurement System (MPMS) in between 2 K and 300 K. A QD VersaLab Vibrating Sample Magnetometer (VSM) was used to measure the magnetic properties between 50 K and 400 K. All the magnetization data were measured with magnetic field parallel to [001] direction. The internal magnetization field ($H_{int}$)  was determined as $H_{int} = H_{applied} - N*M$  to determine the Curie temperature using Arrott plots~\cite{Arrott1967}. Here, \textit{N} is the demagnetization factor related to sample geometry and \textit{M} is the magnetization. The details of the demagnetization factors determination can be found elsewhere~\cite{Aharoni1998, Lamichhane2015,HfZrMnPTej}. Electrical transport property was measured using the four probe method with a Linear research Inc. meter bridge LR 700 (1mA; 17 Hz excitation). Thin platinum wires were attached to the resistance bar using DuPont 4929N silver paint to make electrical contacts. A MPMS was used as a temperature controller for the electrical transport measurements. The specific heat capacity was measured in a QD physical property measurement system using the relaxation technique.

\section{Results and discussion}

Ce$_{3-x}$Mg$_x$Co$_9$ forms in the rhombohedral structure for all \textit{x} examined. Fig.~\ref{Latticeparametersfinal} shows the variation of the lattice parameters\textit{ a} and \textit{c} with Mg content. The \textit{c} lattice parameter decreases monotonically whereas the \textit{a} lattice parameter is more or less constant up to $x\leq 0.50$.

Figure~\ref{nonmag} shows the zero field cooled (ZFC) temperature dependent magnetization \textit{M(T)} for $x\leq 0.50$. Although CeCo$_3$ was identified as a Pauli paramagnetic compound long ago~\cite{Buschow1980JLCM}, there has been some room for question because of the presence of a low temperature upturn in temperature dependent magnetization~\cite{TejCeCo32018}. Moreover recent density functional calculation showed CeCo$_3$ could order ferromagnetically at low temperature~\cite{PandeyCeCo32018}. With $x\leq 0.24$ of Mg addition, the low temperature magnetization remains temperature independent and manifests Pauli paramagnetism as shown in the inset of Fig.~\ref{nonmag}. These temperature independent magnetization data for \textit{x} = 0.01, 0.16, and 0.24 confirm that for these \textit{x}-values, Ce$_{3-x}$Mg$_x$Co$_9$ is Pauli paramagnetic system. Since \textit{x} = 0.01 sample is Pauli paramagnetic down to 2 K, this suggests that pure CeCo$_3$ may also be a Pauli paramagnetic. The low temperature upturn~\cite{TejCeCo32018} is most likely  associated with magnetism of impurity ions or traces of  extrinsic magnetic impurity. 

As we increase the value of x, a low temperature upturn starts to become visible for \textit{x} = 0.35 suggesting it may be close to a critical concentration for achieving the quantum phase transition. For x = 0.43 and 0.50, the upturns in \textit{M(T)} data upon cooling develop large enough magnetization to suggest that they are ferromagnetic samples. The kink only visible on the ZFC \textit{M(T)} data for \textit{x} = 0.43 and 0.50 could be related to reorientation of ferromagnetic domains near the transition temperature.

\begin{figure}[!h]
\centering
\includegraphics[scale =0.35]{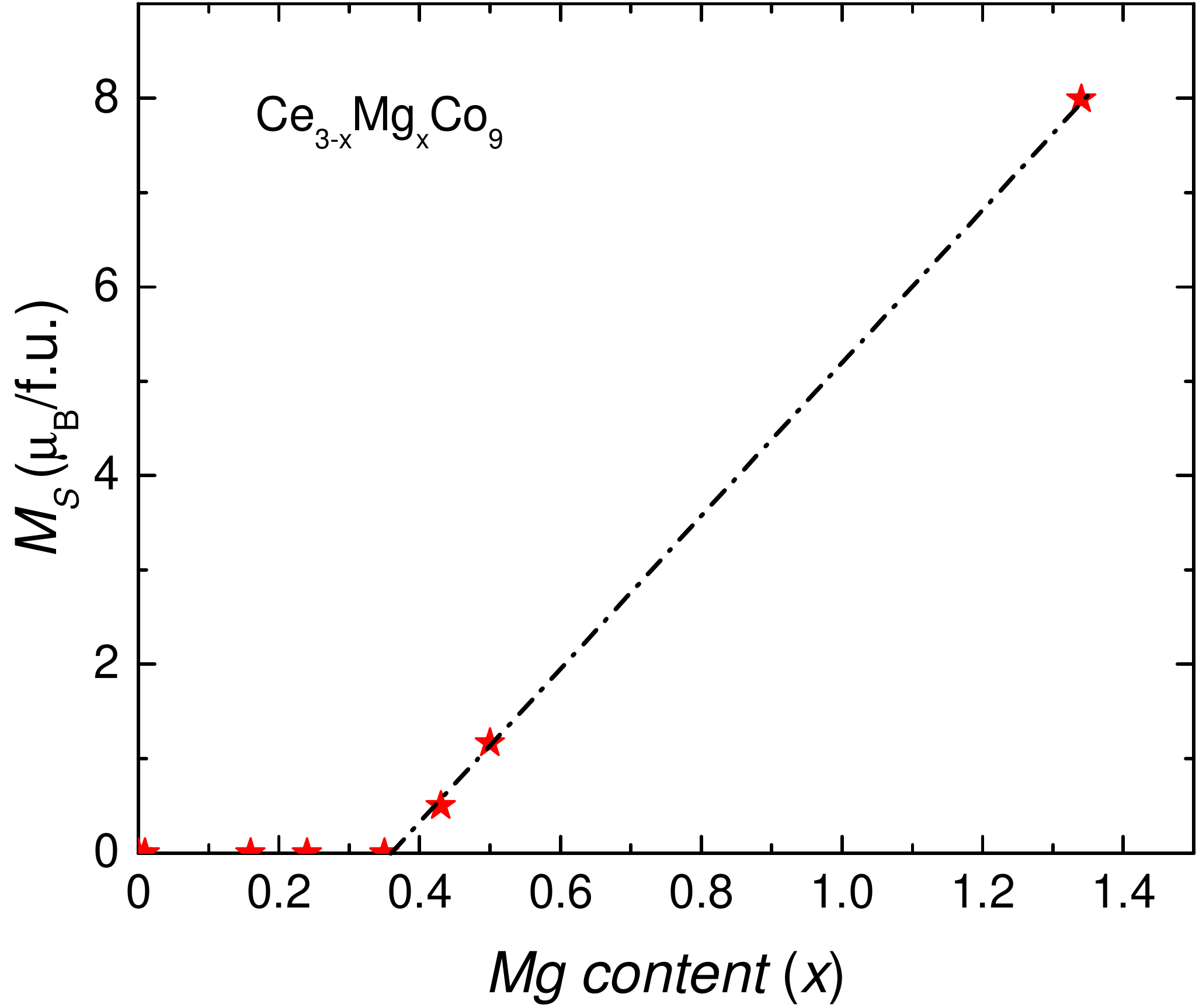}
\caption{\textit{M}$_S$ - \textit{x} relation of Ce$_{3-x}$Mg$_x$Co$_9$ samples. The graph includes spontaneous magnetization \textit{M}$_S$ only for single crystalline samples estimated from the Y-intercept of the linear fit of high field 2 K \textit{M(H)} data along the easy axis [001]. The dash-dot line is a guide for the eyes.}
\label{MSx}
\end{figure}

\begin{figure}[!h]
\centering
\includegraphics[scale =0.35]{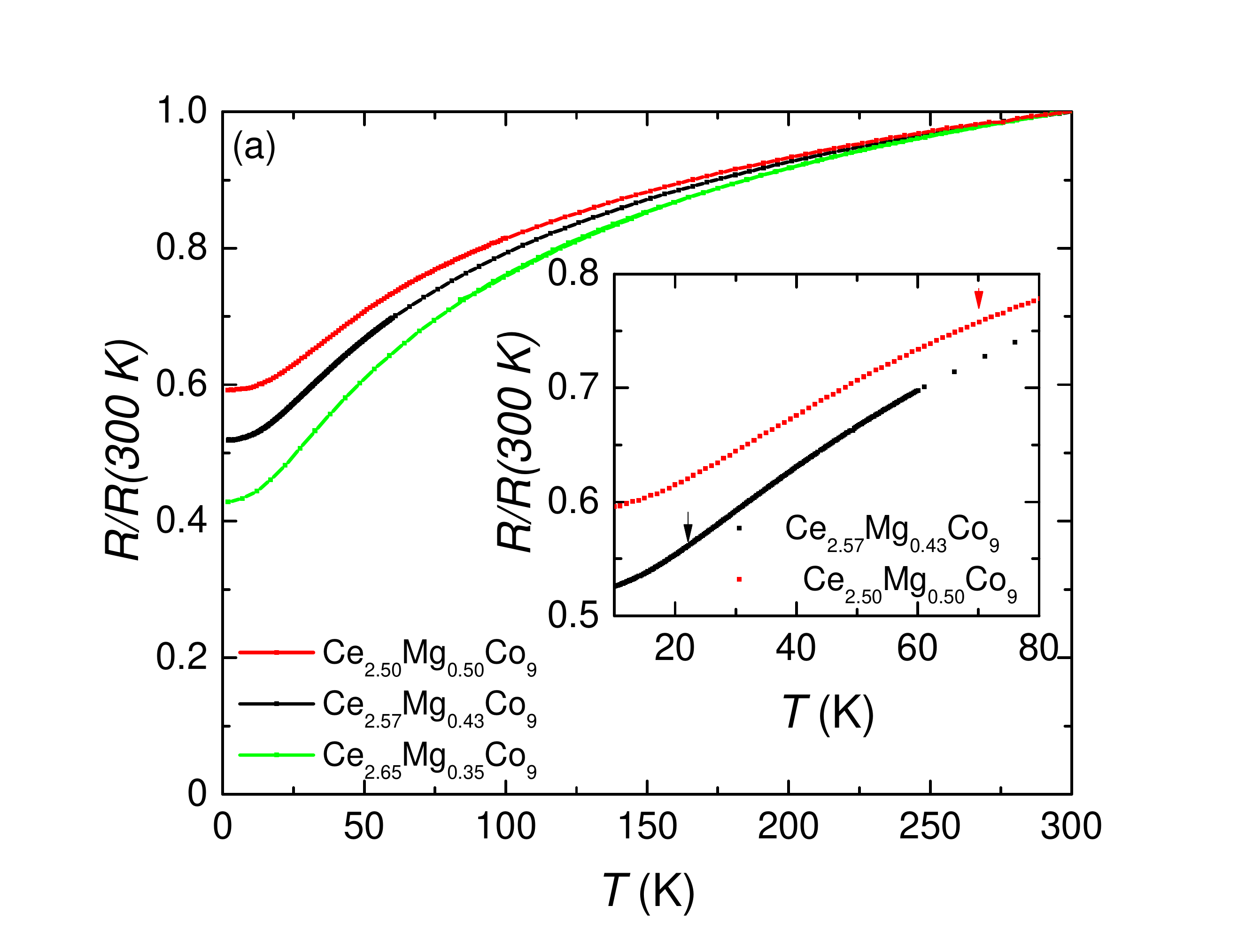}
\caption{Normalised resistance ($\frac{R(T)}{R(300~{\rm K})}$)  data for Ce$_{3-x}$Mg$_x$Co$_9$ samples near the critical concentration: \textit{x}$\sim$0.35, 0.43 and 0.50. No anomalies were observed in any of the measured resistances. The inset shows the enlarged high data density \textit{R(T)} measurements near Curie temperature inferred from Arrott plots indicated with corresponding arrows for \textit{x} = 0.43 and \textit{x} = 0.50 samples.}
\label{NormRT}
\end{figure}

\begin{figure}[!h]
\centering
\includegraphics[scale =0.3]{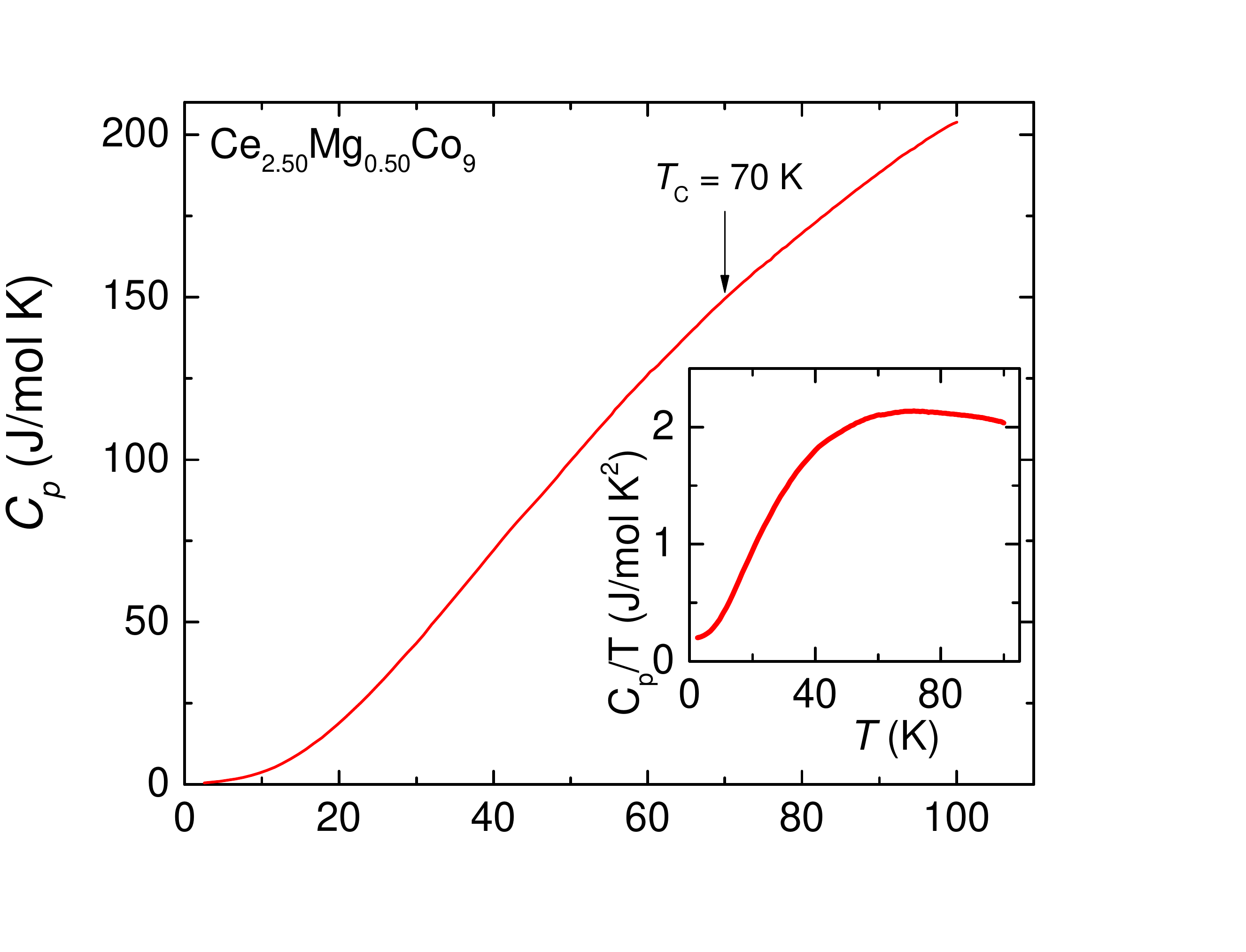}
\caption{Constant pressure specific heat capacity of Ce$_{2.50}$Mg$_{0.50}$Co$_9$ sample. The inset shows the $\frac{C_p}{T}$ as a function of temperature. No clear signature of the ferromagnetic phase transition was observed in either C$_p$ or $\frac{C_p}{T}$ around 70 K (pointed with downward arrow).}
\label{Specificcheatcom}
\end{figure}

To better understand the evolution of ferromagnetism with Mg content \textit{x}, a detailed analysis of easy axis \textit{M(H)} data was performed for all samples. Figure~\ref{MH-nonmag} shows the 2 K \textit{M(H)} data for the non-ferromagnetic samples with EDS inferred Mg concentrations \textit{x} = 0.01, 0.16, 0.24 and 0.35. In Fig.~\ref{TL922-MH},  for the Ce$_{2.65}$Mg$_{0.35}$Co$_9$ sample, we can see an increasing low-field induced magnetization upon cooling indicating that it may be getting close to a ferromagnetic transition.  The lower right inset in Fig.~\ref{TL922-MH} shows ZFC and field cooled (FC) \textit{M(T)} data at 300 Oe applied field. This reversible nature of the \textit{M(T)} data argues against ferromagnetic transition for $T>2.0$ K.

In order to study the \textit{x} = 0.35, 0.43 and 0.50 samples in greater detail, \textit{ M(H)} loops were performed as shown in Figs.~\ref{TL922-MH},~\ref{MH-TL923}, and~\ref{MH-TL924} along with the determination of Curie temperature using Arrott plot method as shown in Figs.~\ref{TL922Arrott},~\ref{TL923Arrott} and~\ref{TL924Arrott}. Within the framework of Arrott plot analysis, straight Arrott curves through the origin suggest a mean field interaction in the magnetic system and identify the Curie temperature.  The Arrott plot of \textit{x} = 0.35 sample does not manifest straight lines that go through origin (as shown in Fig.~\ref{TL922Arrott}). This means that the Ce$_{2.65}$Mg$_{0.35}$Co$_9$ is non-ferromagnetic down to 2 K despite the slight upturn in \textit{M(T)} data.  On the other hand, the Arrott plot data become straighter for Ce$_{2.57}$Mg$_{0.43}$Co$_9$  (Fig.~\ref{TL923Arrott}(b)) and almost an ideal mean-field-like for Ce$_{2.50}$Mg$_{0.50}$Co$_9$ (Fig.~\ref{TL924Arrott}). From these analyses we can infer $T_{\rm C} \approx$ 25 K, 70 K for \textit{x} = 0.43 and 0.50 respectively.

The 2 K  \textit{M(H)} data for \textit{x} = 0.43 and 0.50 samples do not saturate up to 7 T applied field in these experiments, as shown in Figs.~\ref{MH-TL923} and~\ref{MH-TL924}. However, higher magnesium containing samples e.g. \textit{x}= 1.34 were well saturated with 3 T applied field in our previous work~\cite{TejCeCo32018}. This could be the evidence that the doping induced magnetism is more itinerant for lower Mg content and becomes more local moment-like with higher content of Mg.

Figures~\ref{xTC-Phasediagram} and~\ref{MSx} present the $T_{\rm C} - x$ and $M_{\rm S} - x$ phase diagrams for the Ce$_{3 - x}$Mg$_x$Co$_9$ system. Both figures identify $0.35 < x <0.40$ as the critical concentration region for the quantum phase transition from a Pauli  paramagnet to ferromagnetic state.

To further analyze the nature of the phase diagram and the quantum phase transition region, electrical transport properties of the samples around the critical composition  and specific heat capacity of the ferromagnetic, \textit{x} = 0.50 sample, were studied. Figure~\ref{NormRT} shows the normalized resistance of  Ce$_{3-x}$Mg$_x$Co$_9$ single crystalline samples for \textit{x} = 0.35, 0.43 and 0.50. Figure~\ref{Specificcheatcom} shows the temperature dependent specific heat data for \textit{x}= 0.50. Neither resistance nor specific heat data manifest clear signatures of ferromagnetic transitions. In the Stoner model, $\Delta C = \frac{M_0^2}{\chi_0T_C}$ would give a discontinuity around 2 Jmol$^{-1}$K$^{-1}$ at $T_C$ for $x = 0.5$, but if spin fluctuation exist above $T_C$ then discontinuity will be smaller~\cite{Mohn1989}. For the resistance data the anticipated loss of spin disorder features may be obscured by the clearly large and increasing disorder scattering. Additionally, doping disorder could have broadened the ferromagnetic transition so that transition feature was undetectable in resistance and specific heat measurements. The lack of a clear feature in specific heat for \textit{x} = 0.50 suggests that there may be relatively little entropy loss associated with the transition. This is consistent with the small spontaneous moment (Fig.~\ref{MSx}) as well as with this system being a fragile, itinerant moment, ferromagnet.

\section{Conclusions}
The temperature dependent magnetic, electrical transport properties and specific heat capacity of the low Mg content doped Ce$_{3-x}$Mg$_x$Co$_9$ samples were studied using flux grown single crystalline samples. From the \textit{T}$_C$-\textit{x} phase diagram, the critical concentration for quantum transition between Pauli paramagnetic and ferromagnetic ground states is determined to be $0.35 \leq x \leq 0.40$. 


\section{Acknowledgements}

Dr. Warren Straszheim is acknowledged
for doing SEM on various samples. This research was supported by the Critical Materials Institute,
an Energy Innovation Hub funded by the U.S. Department
of Energy, Office of Energy Efficiency and Renewable Energy,
Advanced Manufacturing Office. This work was performed at the Ames
Laboratory, operated for DOE by Iowa State University under
Contract No. DE-AC02-07CH11358.
\bibliographystyle{apsrev4-1}

\end{document}